\documentclass[%
aps,
prapplied,
reprint,
citeautoscript,
superscriptaddress,
 amsmath,amssymb,
]{revtex4-1}

\usepackage{graphicx}
\usepackage{dcolumn}
\usepackage{bm}
\usepackage{comment}
\usepackage[normalem]{ulem}

\DeclareMathOperator{\sech}{sech}

\usepackage[dvipsnames]{xcolor}

\begin{document}

\preprint{APS/123-QED}

\title[]{Antiferromagnetic Bloch line driven by spin current \\as room-temperature analog of a fluxon in a long Josephson junction }
\author{R.V. Ovcharov}
\affiliation{ 
Department of Physics, University of Gothenburg, Gothenburg 41296, Sweden
}

\author{B.A. Ivanov}
\affiliation{
Institute of Magnetism of NASU and MESU, Kyiv 03142, Ukraine
}
\affiliation{
Radboud University, Institute for Molecules and Materials, Nijmegen 6525 AJ, Netherlands
}

\author{J. \AA kerman}
\affiliation{ 
Department of Physics, University of Gothenburg, Gothenburg 41296, Sweden
}

\author{R. S.  Khymyn}%
\affiliation{ 
Department of Physics, University of Gothenburg, Gothenburg 41296, Sweden
}

\date{\today}

\begin{abstract}
Antiferromagnets (AFMs) are promising materials for future high-frequency field-free spintronic applications. Self-localized spin structures can enhance their capabilities and introduce new functionalities to AFM-based devices. Here we consider a domain wall (DW), a topological soliton that bridges a connection between two ground states, similar to a Josephson junction link between two superconductors. We demonstrate the similarities between DWs in bi-axial AFM with easy-axis primary anisotropy, driven by a spin current, and long Josephson junctions (LJJs). We found that the Bloch line (BL) in DWs resembles the fluxon state of Josephson junctions, creating a close analogy between the two systems. We propose a scheme that allows us to create, move, read, and delete such BLs. This transmission line operates at room temperature and can be dynamically reconfigured in contrast to superconductors. Results of a developed model were confirmed by micromagnetic simulations for Cr$_2$O$_3$ and DyFeO$_3$, i.e., correspondingly with weak and strong in-plane anisotropy. Overall, the proposed scheme has significant potential for use in magnetic memory and logic devices.

\end{abstract}

\maketitle

\section{Introduction}
Most spintronic nano-devices nowadays are based on ferromagnetic materials (FMs), which have undisputed benefits, such as the tunability of their properties (e.g., operational frequency) by an external applied field. Also, spin dynamics in FMs is sensitive to the demagnetizing
fields, so one can tune the desired spin mode profile by the corresponding shaping of the sample. For example, such a method can be employed to maximize the overlap between oscillating mode and maximum current density in active spintronic devices, known as spin-Hall and spin-transfer-torque nano-oscillators \cite{dvornik2018origin, mazraati2018auto}. The combination of the above two methods results in a broad variety of spin dynamics in identical stacks of thin films, simplifying the production technology and enriching the areas of possible applications  \cite{chen2016spin}.

Antiferromagnets (AFMs) raise interest for future spintronic applications thanks to their intriguing properties, such as intrinsic ultra-fast spin dynamics without the necessity of an applied magnetic field and the absence of stray fields, which pave the way to the compact spintronic devices operating in the sub-THz frequency range. However, the absence of tunability by an applied field and shape anisotropy makes AFM-based devices more dependent on inherent spatial spin distributions and the material magnetic parameters. Particularly, the localization of spin dynamics in AFMs is hardly achievable by an effective field profile. It was proposed instead to employ different spin structures, such as domain walls (DWs), skyrmions, and other solitons since they can exhibit rich dynamics and hence embroad the variety of spintronic devices  \cite{gomonay2018antiferromagnetic}.

 \begin{figure}[ht!]
\includegraphics[width=\linewidth]{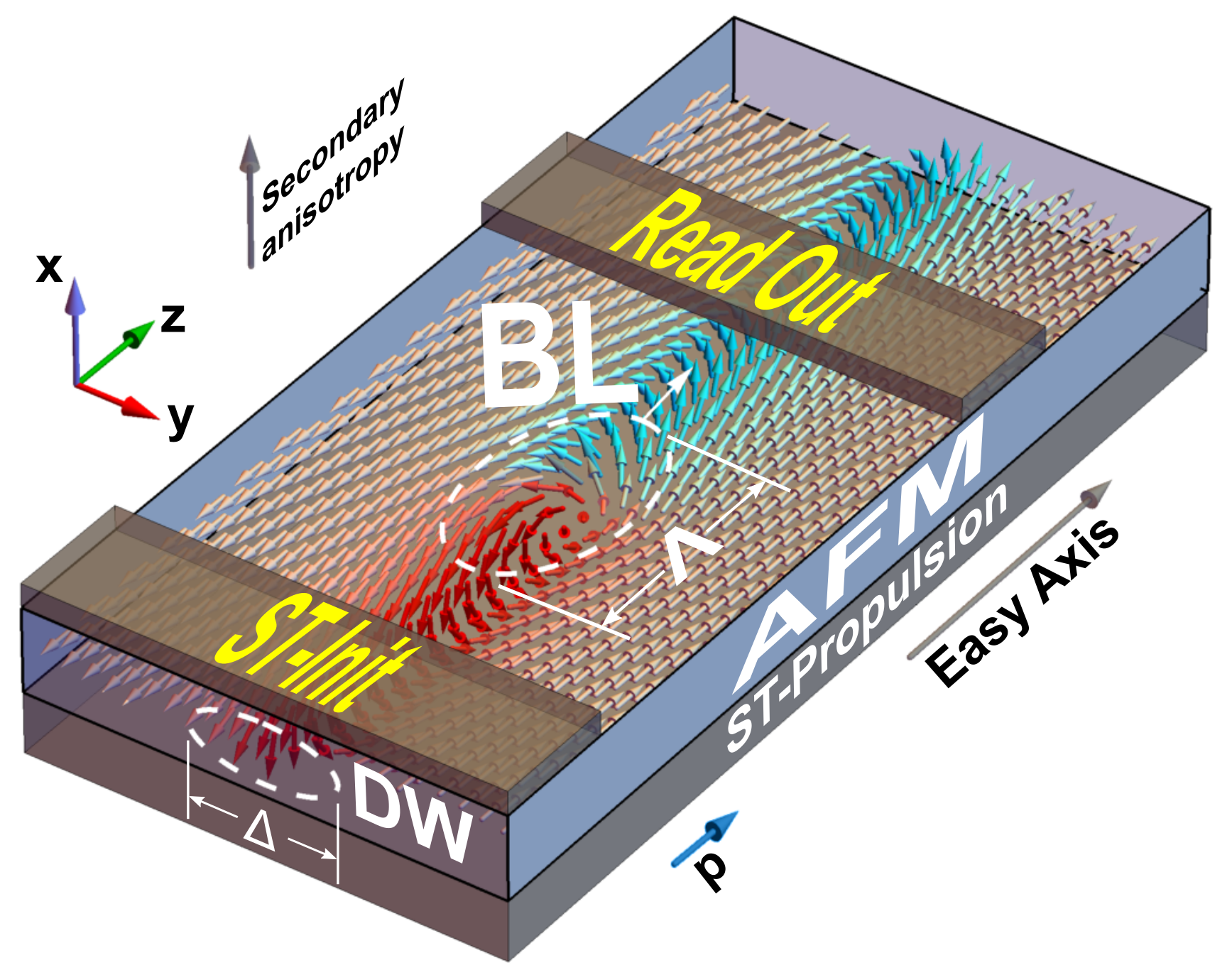}
 \caption{Schematic diagram illustrating a transmission line on a DW with a BL as an information carrier. The creation of the BL is achieved by applying a short current pulse to the ST-Init source, after which the ST-Propulsion source supports the movement of the BL. The readout gate captures the response from the passage of the BL via the spin-pumping mechanism. Gray arrows show the directions of the easy axis and secondary anisotropy, and the blue arrow indicates the direction of the spin current polarization $p$.}
\label{img:schema}
\end{figure}

Magnetic anisotropy defines not only the resonant properties of the AFM but also the trajectories of driven spin dynamics. AFM magnetic ordering is inherent for materials with a broad variety of crystal structures and correspondingly a wide range of anisotropy symmetries, which quite often can be reduced to the case of a bi-axial anisotropy with orthogonal axes. The sign of a primary anisotropy constant defines the predominant behavior of a magnetic, which can be formally divided into easy-plane type or easy-axis. Easy-plane AFMs are of special interest for spin-current driven devices due to the low threshold current density of the spin dynamic excitation, which is defined by the secondary anisotropy. Even more important, that easy-plane magnets provide the topology necessary for spin superfluidity, where the angle of the magnetic order parameter in easy-plane corresponds to the phase of the macroscopic wave function in actual superfluids or superconductors  \cite{sonin2010spin, takei2014superfluid}. The presence of the second weak anisotropy brakes the necessary symmetry, but in this case, an AFM spin-torque nano-oscillator is an analog of Josephson junction\cite{khymyn_antiferromagnetic_2017, barone1982physics}, while a magnetic strip subjected to the spin injection resembles a long Josephson junction (LJJ) and DWs in easy-plane magnets correspond to phase vortices  \cite{hill2018spin}. 

A DW in easy-axis FMs also can be a host of spin superfluid, which can be employed for skyrmion generation and detection  \cite{kim2017magnetic}. Thus, here we consider a domain wall in bi-axial AFM of easy-axis type and a Bloch line (BL) in it driven by a spin current. Thanks to their topological stability and nanoscale size, BLs are promising candidates for information carriers. The idea of using the BL in FMs as a memory ``bit'' controlled by an external magnetic field arose long ago  \cite{konishi1983newultradensity, humphrey1985vertical}. However, the development of spintronics has revived interest in the study of BLs,  \cite{herranen2015domain, yang_racetrack_2021} particularly due to the advances in DW creation and manipulation. Thus, in magnetoelectric AFMs like Cr$_2$O$_3$, a DW can be created using electromagnetic field cooling \cite{hedrich2021nanoscale}. In AFM orthoferrites (e.g., DyFeO$_3$), DWs can be formed by applying a nonuniform magnetic field in the canted (weak ferromagnetic) state \cite{bar1985dynamics} or by a combination of magnetic field and stress in a ``pure'' AFM state \cite{afanasiev2016control}.

We show that an AFM BL driven by a spin current is a close analog of a Josephson phase vortex in an LJJ, known as fluxon. We propose a scheme that allows to create, move, read, and clear such BLs. Compared to superconductors, this transmission line operates at room temperature and can be dynamically reconfigured.

\section{Model and problem formulation}
We consider a two-sublattice antiferromagnet where the magnetization of the sublattices ($\mathbf{M}_1$ and $\mathbf{M}_2$, respectively) are equal in magnitude and opposite in direction. For the description of the AFM it is convenient to introduce normalized antiferromagnetic  N\'eel $\mathbf{l}=(\mathbf{M}_1 - \mathbf{M}_2)/M_s$ and net magnetization $\mathbf{m}=(\mathbf{M}_1 + \mathbf{M}_2)/M_s$ orthogonal vectors, where $M_s$ is the saturation magnetization at a parallel orientation of sublattices. In the case of the dominant exchange interaction, when the representative exchange field $H_{ex}$ that aligns magnetic moments of sublattices in opposite directions is significantly large, the net magnetization is small $|\mathbf{m}|\ll 1$, and $\mathbf{l}$ can be considered a unit vector $|\mathbf{l}| \approx 1$. This simplification allows the description of the low energy dynamics in AFMs by a closed-form equation with a single variable $\mathbf{l}$, known as the $\sigma$-model equation \cite{baryakhtar_dynamics_1994}. Parametrizing the N\'eel vector in spherical angular coordinates $\mathbf{l} = \{ \sin \theta \cos \phi, \sin \theta \sin \phi, \cos \theta\}$ the Lagrangian density of the $\sigma$-model in the nondissipative limit can be written as follow  \cite{kosevich_magnetic_1990, gomonay_spin_2010}:
\begin{equation}
 \mathcal{L}=\frac{M_s}{2\gamma \omega_{ex}}\left[\dot \theta^2 -c^2 \theta'^2+\sin^2 \theta (\dot \phi^2 - c^2\phi'^2)\right]- w_a(\theta, \phi) ,
 \label{eq:lagrangian}
 \end{equation}
 where an upper dot and prime denote time and space derivatives, respectively, $\gamma$ is a gyromagnetic ratio, $\omega_{ex}=\gamma H_{ex}$ is the frequency defined by the exchange field, $c=\gamma \sqrt{H_{ex}A/M_s}$ is the characteristic speed of magnons, where $A$ is the exchange stiffness. The last term $w_a$ describes the anisotropy energy of the AFM and is taken in the form:
 \begin{equation}
     w_a(\theta, \phi) = \frac{M_s}{2 \gamma \omega_{ex}}(\omega_0^2 + \omega_{b}^2\sin^2\phi)\sin^2\theta,
 \end{equation}
 where the first term $\omega_0 = \gamma \sqrt{H_{ex} K / M_s}$ defines primary uniaxial anisotropy $K$ of the easy-axis type ($K>0$), and the second term defines the simplest form of anisotropy in the basal plane $K_{b}$ for an AFM with an easy axis of the second order $C_2$.
 
 The contribution of nonconservative processes - energy loss due to the dissipation combined with the torque-induced energy gain, is taken into account by choosing the dissipative Rayleigh function in the form  \cite{gomonay_spin_2010, ivanov_spin_2020}:
 
\begin{eqnarray}
\mathcal{R}=\frac{ M_s }{2\gamma}\left[\alpha (\dot \theta^2+\dot\phi^2 \sin^2 \theta ) - 2 \tau \dot\phi\sin^2 \theta \right],
\label{eq:dissipative}
\end{eqnarray}
where $\alpha$ is an effective Gilbert damping, $\tau$ is the spin torque (ST) amplitude expressed in frequency units. Here, we utilize the spin-Hall effect for the ST generation, see Fig. 1; a direct current passing through the heavy metal layer with spin-orbit interaction (e.g., Pt) injects a polarized spin current into the AFM layer. This injection gives rise to a nonconservative ST, so-called spin-orbit torque, $\tau = \sigma j$, where $j$ is the density of the electrical current, and $\sigma$ is the ST efficiency. Alternatively, spin-transfer torque (produced by current polarization through the ferromagnetic layer) can also be employed. The torque polarization is chosen to be along the AFM easy axis.

The Lagrangian of the form (\ref{eq:lagrangian}) in the limit of purely uniaxial symmetry specifies solution for a static DW with $\cos \theta = \tanh (y/\Delta)$ and $\phi = \phi_0$, where $\Delta = \sqrt{A/K}$ is the DW thickness and $\phi_0$ determines the rotation plane of the $\mathbf{l}$ vector in the wall. A fairly detailed analysis of the DW dynamics can be provided by representing DW as a particle-like object with two degrees of freedom. Thus, substitutions $y \rightarrow y_0 - Y(t)$ and $\varphi_0 \rightarrow \Phi(t)$, where $Y$ describes the displacement of the DW center from the initial position $y_0$, and $\Phi$ is the DW phase, define translational and rotational degrees of freedom respectively. This description refers to the so-called collective coordinates approach, which is used to study the DW forced dynamics  \cite{clarke_dynamics_2008, kim_propulsion_2014, ovcharov_spin_2022}. This method can be easily extended to study dynamics along the DW by considering an additional dependence of collective variables on the coordinate $z$ \cite{khodenkov_bloch_2006}.

In the subsequent discussion, we focus on the DW phase dynamics; the dissipative function (\ref{eq:dissipative}) does not include terms with ST polarized perpendicularly to the easy axis, as they lead to the movement of the DW \cite{shiino_antiferromagnetic_2016, ovcharov_spin_2022}. The fluctuation of the DW position due to imperfect polarization can be removed by the DW pinning (for example, by employing the nanoconstriction \cite{ovcharov_spin_2022}), so we assume that the domain wall is placed along the straight line defined by $Y(z, t) = const$. The width of the domain wall in the presence of the basal plane anisotropy depends on its phase $\Delta = \sqrt{A/(K+K_b\sin^2 \Phi)}$. We neglect the effect of the dynamic response of the DW width by assuming, in the limit of small secondary anisotropy $K_b/K \ll 1$, that it remains constant. It allows us to obtain the effective Lagrangian and energy dissipation function by integrating the Lagrangian density (\ref{eq:lagrangian}) and dissipative function (\ref{eq:dissipative}) over $y$ in infinite limits with parametrization of the DW profile:

\begin{eqnarray}
    \mathcal{L}_{\Phi}=\frac{\Delta M_s}{\gamma \omega_{ex}} \left[ \dot{\Phi}^2 - c^2 \Phi^{\prime 2} -\omega_b^2 \sin^2 \Phi \right],\label{eq:Leff}\\
    \frac{dE}{dt}= \dot \Phi \left[ -\frac{2 \Delta M_s }{\gamma}\alpha \dot \Phi +  \frac{2 \Delta M_s }{\gamma}\tau \right]. \label{eq:dEeff}
\end{eqnarray}

 The term in brackets in the energy dissipation rate function (\ref{eq:dEeff}) indicates the nonconservative moment of force $F_{\Phi}=\dot E / \dot \Phi$ associated with DW rotational dynamics that we incorporate into a standard Euler-Lagrange equation. In this way, the effective dynamic equation for the collective variable $\Phi$ takes the form of the perturbed sine-Gordon equation \cite{barone_theory_1971, ustinov_solitons_1998, baryakhtar_dynamics_1994}:

\begin{equation}
  2\omega_{b}^{-2} \ddot \Phi   - 2 \Lambda^2  \Phi^{''}   +  \sin  2\Phi = \tau / \tau_{c} -  2\omega_{\alpha}^{-1} \dot \Phi, \label{eq:mainPhi}
\end{equation}
where characteristic length and time scales are given by the intrawall magnetic length $\Lambda = \sqrt{A/K_{b}}$ and frequency of magnon gap $\omega_{b} = c/\Lambda$, damping rate is inversely proportional to the Gilbert constant $\omega_{\alpha}=\omega_{b}^2/\alpha \omega_{ex}$ and $\tau_{c} = \omega_{b}^2/2\omega_{ex} $ is the critical torque needed to overcome the anisotropy barrier.
This equation is analogous to the equation that describes the dynamics of the Josephson phase $\Phi$, the phase difference of the wave function of superconducting electrons, in a long Josephson junction  \cite{ustinov_solitons_1998, baryakhtar_dynamics_1994}. Here applied spin-transfer torque represents a bias current through the junction, anisotropy in a basal plane defines Josephson plasma frequency $\omega_p$, and characteristic speed of magnons, defined by nonuniform exchange energy, corresponds to the Swihart velocity $c_0 = \lambda_J \omega_p$, where $\lambda_J$ is the Josephson penetration depth. Comparing the characteristic values of these systems with the same time scale, junctions` spatial scale \cite{fedorov_fluxon_2014, scott_propagation_1970} ($\lambda_J$ is of the order of 10 to 500 microns) is much larger than the intrawall characteristic length ($\Lambda$ is of the order of tens of nanometers). It makes the magnetic system a suitable candidate for more compact devices that, in addition, can operate without the need for cryogenic temperatures.

If the length of the system $L$ is significantly larger than a characteristic length $\Lambda(\lambda_J)$, Eq. (\ref{eq:mainPhi}) admits a solitonic solution of the kink type. The following soliton in an LJJ system corresponds to a Josephson supercurrent vortex, often called a fluxon, since it carries a single quantum of magnetic flux $\Phi_0 = h /2 e$. The equivalent kink-type solution also exists in the proposed AFM DW-based system and corresponds to the Bloch line - the spatial transition of spins that smoothly connects two ground states of the DW. It is instructive to consider the static part of the energy $W = c^2 \Phi^{'2} + \omega_{b}^2\sin^2 \Phi$, see Lagrangian (\ref{eq:Leff}),  to examine the characteristics of this spin structure. The equilibrium values of the static homogeneous DW phase are found by minimizing the intrawall anisotropy energy and correspond to $\Phi = 0, \pi$. This discrete degeneracy of the DW energy accounts for the existence of a self-localized BL inside the DW associated with spontaneously broken symmetry. The structure of the BL profile originates from the interplay between inhomogeneous exchange and anisotropy terms, which together with boundary conditions $\Phi(-\infty) = 0, \Phi(+\infty)=\pi$ is described by the same with DW soliton profile: $\cos \Phi = \tanh (z / \Lambda)$. The length of the BL $\Lambda$ is greater than the DW width $\Delta$ as it is determined by secondary anisotropy $\Lambda/\Delta = \sqrt{K/K_{b}}$.

\section{Bloch line dynamics}
\subsection{Simple perturbation theory}

We now proceed to a study of the dynamic properties of the BL by a simple perturbation theory, but as we will see from the results of micro-magnetic simulations, it will require substantial modification. The left-hand side of the phase dynamics equation (\ref{eq:mainPhi}) shows ``Lorentz''-invariant behavior with $c=\omega_{b}\Lambda$ as a limiting velocity. By virtue of this property, the solution describing a moving BL can be obtained by simple Lorentz transformations $z \rightarrow vt$ and $\Lambda \rightarrow \Lambda_0\sqrt{1-(v/c)^2}$ applied to the static profile. Thus, moving by ``inertia'' in a nondissipative limit, BL behaves like a ``relativistic'' particle, decreasing its width, according to the well-known relativistic effect of the  Lorentz contraction, as the velocity tends to $c$  \cite{kosevich_magnetic_1990}.

In the presence of perturbative terms, namely dissipation that inhibits BL and ST that leads to the BL propulsion, it is impossible to find an exact analytical solution for Eq. (\ref{eq:mainPhi}). However, a rather complete analysis of dynamic properties can be done using the aforementioned adiabatic approximation in collective coordinates \cite{kivshar_dynamics_1989}. Supposing the dissipation rate is significantly low $\omega_{\alpha}^{-1}\omega_{b} \ll 1$ and the amplitude of the applied spin-torque is less than the critical value $\tau < \tau_{c}$, one can assume that the structure of the BL does not change under the action of perturbations. In this case, the rigid soliton solution can be represented as a solid particle of a characteristic size $\Lambda$, with temporal dynamics being held by a collective coordinate $Z$, determining the position of the BL. Thus, using the magnetization distribution for the BL, we can calculate the friction force acting on this particle depending on its velocity. In turn, the energy influx from ST is treated as an external driving force. The corresponding equation of motion takes the form:
\begin{equation}
   \frac{d}{dt}\left(\frac{\dot{Z}}{\Lambda}\right) = \frac{\pi \tau \omega_{ex}}{2}  - \alpha  \omega_{ex} \frac{\dot{Z}}{\Lambda},\label{eq:eqOfMotion}
\end{equation}
where the left-hand side is the time derivative of the BL momentum $P_Z = \dot{Z}/\Lambda(\dot{Z})$, and the right-hand side describes propulsion and friction forces. By equating these forces, i.e., considering the energy loss compensation regime in the steady-state motion $Z = vt$, so $\dot{P_Z} = 0$, the dependency of the BL  velocity on the external excitation can be found as:

\begin{equation}
    v = \frac{\mu \tau c}{\sqrt{(\mu \tau)^2 + c^2}},\label{eq:velocity}
\end{equation}
where $\mu=\pi \Lambda_0 / 2\alpha $ has the sense of the mobility of the BL with respect to $\tau$ at low velocity. At small current values, the velocity has a linear dependence $v = \mu\tau$. With a subsequent increase in current, the velocity is saturated, monotonically tending to the limit c. However, there is one effect that limits the maximal velocity of the BL. As has been shown in Ref. \cite{ovcharov_spin_2022}, if the value of the torque exceeds some threshold torque $\tau_c$, the N\'eel vector within the entire domain wall starts to precess around the easy axis. This effect determines the presence of the maximum
driving force $F_{max} = \pi \tau_{c} \omega_{ex} / 2$; the BL can only reach a certain highest velocity $v_{max}$ before the beginning of this precession. The value of this velocity can be easily found by substituting the expression for the critical torque into Eq. (\ref{eq:velocity}) and can be written in the following form:
\begin{equation}
    v_{max} = \frac{\kappa c}{\sqrt{1+\kappa^2}}, \kappa = \frac{\pi \omega_b}{4 \alpha \omega_{ex}}\label{eq:maxvelocity}.
\end{equation}

Although the maximum velocity of the BL increases with increasing anisotropy,  its mobility is inversely proportional to $\omega_b$.

\begin{figure}[hbt!]
\includegraphics[width=\linewidth]{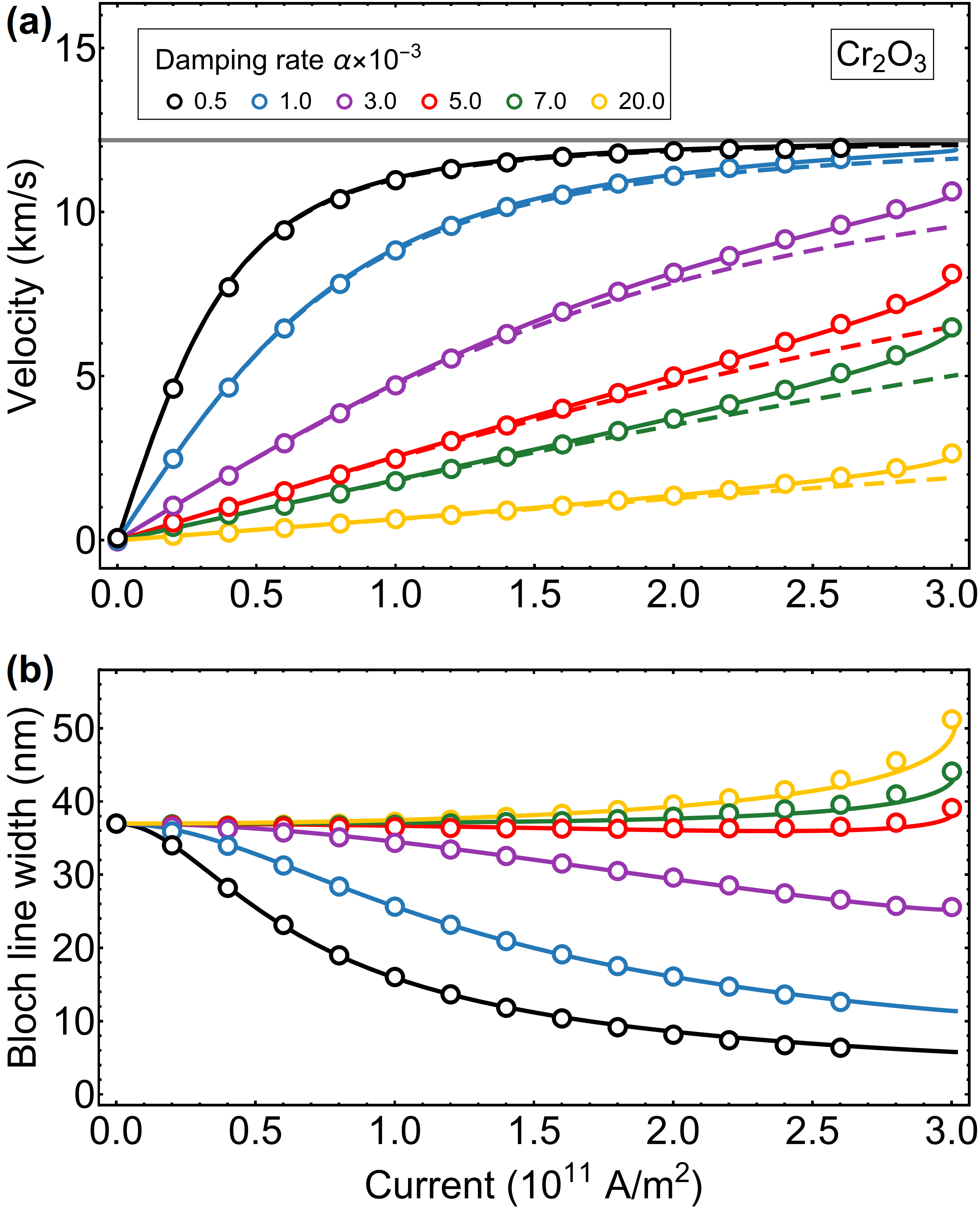}
\caption{(a) The dependence of the Bloch line velocity on current for different values of damping rate in Cr$_2$O$_3$. The circles show the values extracted from the micro-magnetic simulations. Solid and dashed lines are calculated according to the analytical dependence (\ref{eq:velocity}) with and without considering the spin current's effect on the BL profile, respectively. The horizontal gray line shows the limiting speed $c=12$~km/s. (b) The dependence of the Bloch line width on current. The circles show the data of micro-magnetic simulations; the width is found from the derivative of the Bloch line profile at its center. Solid lines are calculated analytically, considering ``vacuum'' level inclination at high currents.  } 
\label{img:blvelocity}
\end{figure}

\begin{figure*}[hbt!]
\includegraphics[width=0.7\linewidth]{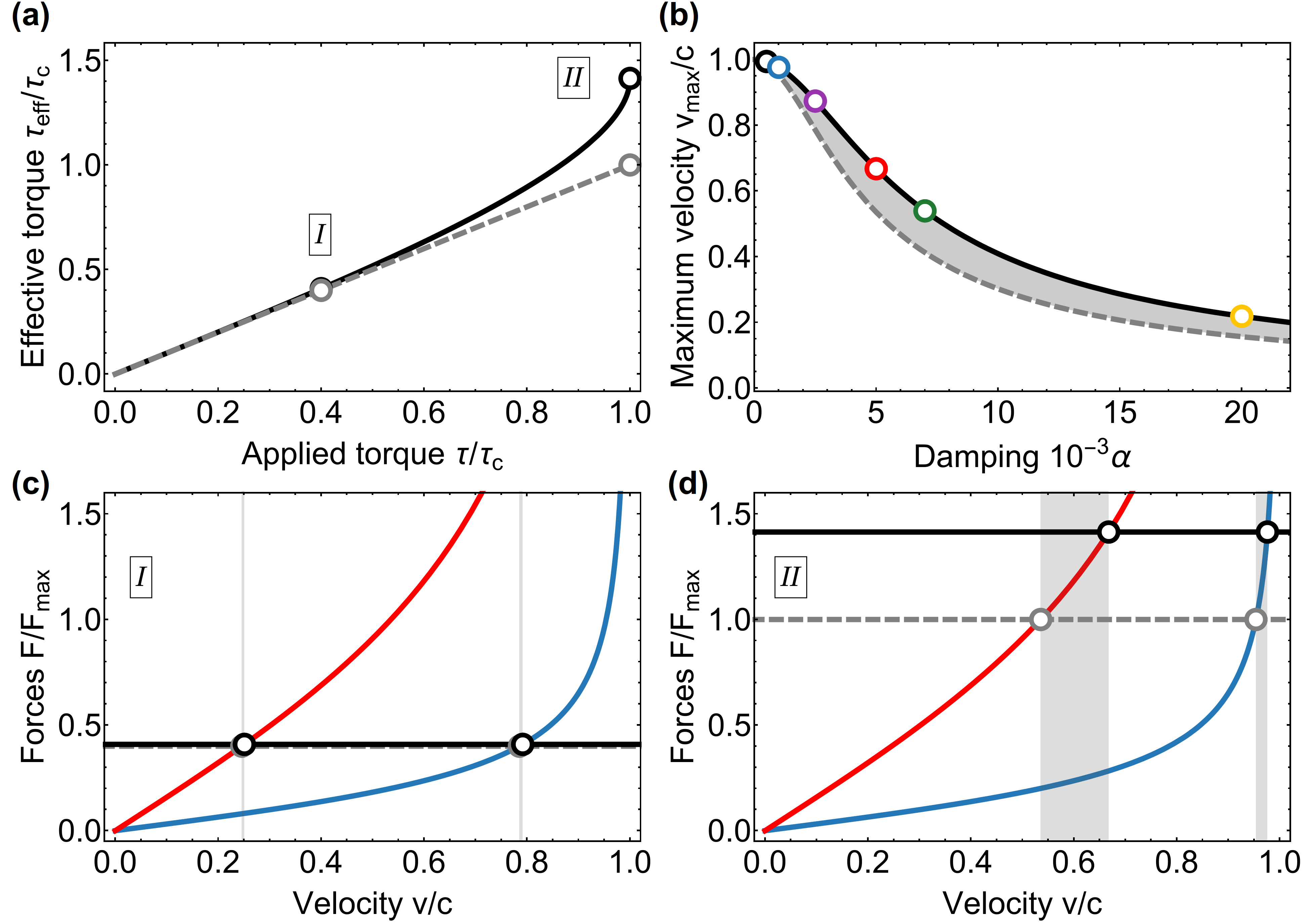}
\caption{
(a) Shows the difference in the effective torque (\ref{eq:efftorque}) acting on BL  (solid black line) from the bias torque (gray dashed line); (b) The dependence of maximum velocity on damping (\ref{eq:maxvelocity}). The solid black line is calculated with the substitution of the maximum effective torque $\tau_{eff}^{max} = \sqrt{2}\tau_{c}$. The circles indicate the dampings used in the simulations for Cr$_2$O$_3$; (c) and (d) The dependences of driving (solid black and dashed gray lines for effective and actual torque correspondingly) and friction (red line is for  $\alpha=5\cdot10^{-3}$, and the blue one is for $\alpha=1\cdot10^{-3}$) forces on the BL velocity for different values of the applied torque shown in subplot (a) by circles. Intersection points determine the resulting velocity. Gray-colored areas indicate differences in velocity calculation with and without considering profile modification presented by Eq. (\ref{eq:profilePhi}).}
\label{img:effective}
\end{figure*}

\subsection{Micro-magnetic simulations}

In order to verify the obtained analytical prediction (\ref{eq:velocity}),  we performed micro-magnetic simulations using MuMax3 solver \cite{vansteenkiste_design_2014} for a system schematically represented in Fig. \ref{img:schema}. The AFM film has lateral sizes 0.25$\times$2.78~$\mu$m$^2$ with a thickness of 5~nm. The method of modeling \cite{clercq_modelling_2017} was described in detail in  \cite{ovcharov_spin_2022}. The selected parameters used for the AFM correspond to the Cr$_2$O$_3$ \cite{foner_high-field_1963, artman_magnetic_1965, li_spin_2020} and given as: $A=5.8$~pJ/m, $H_{ex}=438$~T ($\omega_{ex}/2\pi=12.6$~THz), $M_s = 5.6 \cdot 10^5$~A/m,  the easy axis anisotropy constant along the $z$-axis $K = 40$~kJ/m$^3$ ($\omega_{0}/2\pi=160$~GHz), the value of the weak secondary AFM anisotropy along the $x$-axis $K_{b} = 4$~kJ/m$^3$ ($\omega_{b}/2\pi=50$~GHz). All ST sources utilize the spin-Hall effect in Pt with $\theta_{SH}=0.1$ and are configured to inject spin current polarized along with the easy axis of the AFM. 

The AFM layer is initialized by a centrally-located pre-relaxed DW with a set width of $\Delta=12$~nm. At the initial moment, a short current pulse with an amplitude exceeding the threshold is applied to the edge-localized ST-init source. The twisted DW then relaxes to its ground state, setting a Bloch line with $\Lambda=38$~nm. The pulse parameters for BL creation are determined by the need for phase rotation by $\pi$, e.g., for the case $\alpha=5\cdot10^{-3}$, we applied current $j=7\cdot10^{11}$~A/m$^2$ for the time interval $\Delta t = 12$~ps. The length of the initialization contact $L_{i}$ defines the initial position of the BL and should be longer than its characteristic width $L_{i} \gg \Lambda_0$. 

After BL initialization, a bias current is applied to the ST-propulsion source leading to the movement of the BL. The dependencies of the established BL velocity on the applied current density for different dissipation rates are shown in Fig. \ref{img:blvelocity}. The obtained analytical dependence (\ref{eq:velocity}), shown in Fig. \ref{img:blvelocity} by dashed lines, describes the BL motion at low currents well but differs notably at the currents approaching critical value $j_{c} = \tau_{c}/\sigma = 3\cdot10^{11}$~A/m$^2$.

\subsection{BL beyond the perturbation approach}

To describe the BL dynamics in this region of perturbation parameters, the action of the ST on the BL profile within its frame of reference should be considered. The removal of energy degeneracy is accompanied by the inclination of the phases in ground states. The effect of the ST can be represented with the help of some effective potential \cite{ivanov_spin_2020} $ \Tilde{U}(\Phi) = \sin^2 \Phi - \tau \Phi /\tau_{c}$ . The minimum of this potential sets the value for the ``vacuum'' level $\Psi_0$ that depends on current as $\Psi_0=\arcsin(\tau_{}/\tau_{c})/2$ and reaches its maximum value $\Psi_0 = \pi/4$ at $\tau=\tau_{c}$. Now the profile of the BL can be approximated by the following trial function:

\begin{equation}
\cos \Phi = \cos \Psi_0  \tanh \left( z/\tilde{\Lambda} \right) - \sin \Psi_0 \sech \left(z/\tilde{\Lambda}\right),\label{eq:profilePhi}
\end{equation}
where $\tilde{\Lambda}$ has the sense of an effective BL width. This formula describes well the behavior of $\Phi$ far from the center of the moving BL, the adequacy of this approximation will be checked by comparison with the numerical data.  The value of $\tilde{\Lambda}$ can be estimated as the ratio between exchange stiffness and anisotropy energy at the central point of the inclined BL, i.e., at the point $\Phi = \Psi_0 + \pi/2$. Consequently, one can write the effective width as $\tilde{\Lambda} = \Lambda/\cos \Psi_0$, which means that BL experiences expansion determined by the slope of the vacuum phase $\Psi_0$. This increase in size originates from the modification of the potential energy and should be taken into account when calculating established BL velocity from a power balance, see Fig.~\ref{img:blvelocity}. With the purpose of preserving the physical meaning of mobility $\mu$, it is convenient to rewrite Eq.~(\ref{eq:velocity}) with the use of some effective ST which is defined as:

\begin{equation}
    \tau_{eff} = \tau / \cos \left[ \frac{\arcsin \tau/\tau_{c}}{2} \right].
    \label{eq:efftorque}
\end{equation}

The modification of the BL profile leads to increasing the action of the torque on the BL propulsion. One can imagine that since the N\'eel vector in the ground states away from the BL is ``pumped" $\Phi = \Psi_0$, the energy required to transit from one state to another, i.e., to move the BL at a given speed, is less than in the case when the N\'eel vector should rotate from the state $\Phi = 0$. However, this effect becomes significant only at high torque values (see Fig. \ref{img:effective}a).

Let us discuss the outcome of this effective torque on the BL velocity. As we noted before, the stationary dynamics is determined by the interplay between torque and damping. If the applied torque is small $\tau \ll \tau_c$, the effect is negligible; since $\tau_{eff}\approx\tau$, there is no difference from the perturbation approach. Also, the situation remains unchanged if the material has low damping $\omega_{\alpha}^{-1}\omega_b \ll 1$. The BL velocity quickly goes into the saturation regime where the increase of effective torque compared to the applied one has no impact. The discrepancy with the perturbation theory calculations occurs if the material has moderate damping $\omega_{\alpha}^{-1}\omega_b \sim 1$ and the applied torque approaches the critical value. In this case, the BL velocity does not reach saturation even if the torque is close to critical. Therefore, the higher effective torque action $\tau_{eff}>\tau$ leads to a noticeable increase in velocity, see Fig. \ref{img:effective}.

\begin{figure}[hbt!]
\includegraphics[width=\linewidth]{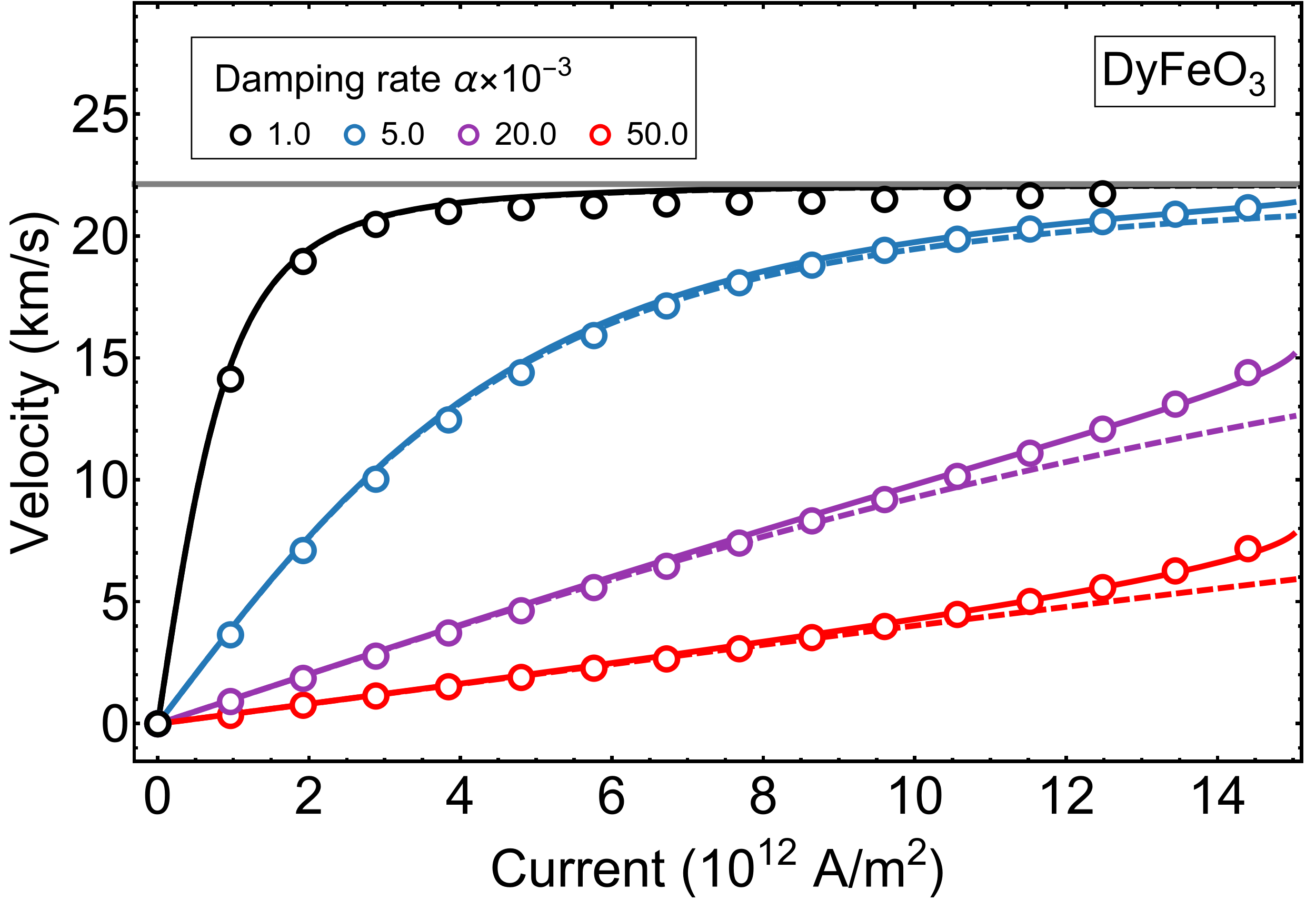}
\caption{The dependence of the Bloch line velocity on current for different values of damping rate in DyFeO$_3$. Circles show the values extracted from the micro-magnetic simulations. Solid and dashed lines are calculated according to the analytical dependence (\ref{eq:velocity}) with and without considering the spin current's effect on the BL profile correspondingly. The horizontal gray line indicates the limiting velocity $c=22$~km/s.}
\label{img:blvelocityDyFeO3}
\end{figure}

As we noted at the beginning, the obtained dynamical equations in collective variables are based on the assumption about the smallness of secondary anisotropy $K_{b}\ll K$, $\omega_b \ll \omega_0$. To proof the adequacy of analytical results for the case of relatively large basal plane anisotropy, we performed additional simulations with dysprosium orthoferrite DyFeO$_3$ as an AFM layer (0.27$\times$1.56~$\mu$m$^2$ in size) using the following parameters \cite{turov_symmetry_2001, iida_spectral_2011, hortensius_coherent_2021}:  $A=18.9$~pJ/m, $H_{ex}=665$~T ($\omega_{ex}/2\pi=19.2$~THz), $M_s = 8.4 \cdot 10^5$~A/m,  the easy axis anisotropy constant $K = 390$~kJ/m$^3$ ($\omega_{0}/2\pi=508$~GHz), the value of the secondary anisotropy $K_{b} = 175$~kJ/m$^3$ ($\omega_{b}/2\pi=340$~GHz). The configuration of ST sources remained unchanged. 

As shown in Fig. \ref{img:blvelocityDyFeO3}, the developed analytical model (\ref{eq:velocity}, \ref{eq:efftorque}) describes the results of micromagnetic simulations quite well. Larger in-plane anisotropy  increases the critical value of current $j_{c} = \omega_b^2/2\sigma\omega_{ex}$ ($15\cdot10^{12}$~A/m$^2$ for DyFeO$_3$) and reduces mobility $\mu \propto \omega_b^{-1}$. However, as noted above, the maximum value of velocity $v_{max}$ (\ref{eq:maxvelocity}) increases, allowing larger damping values to observe the effect.

\begin{figure}[hbt!]
\includegraphics[width=\linewidth]{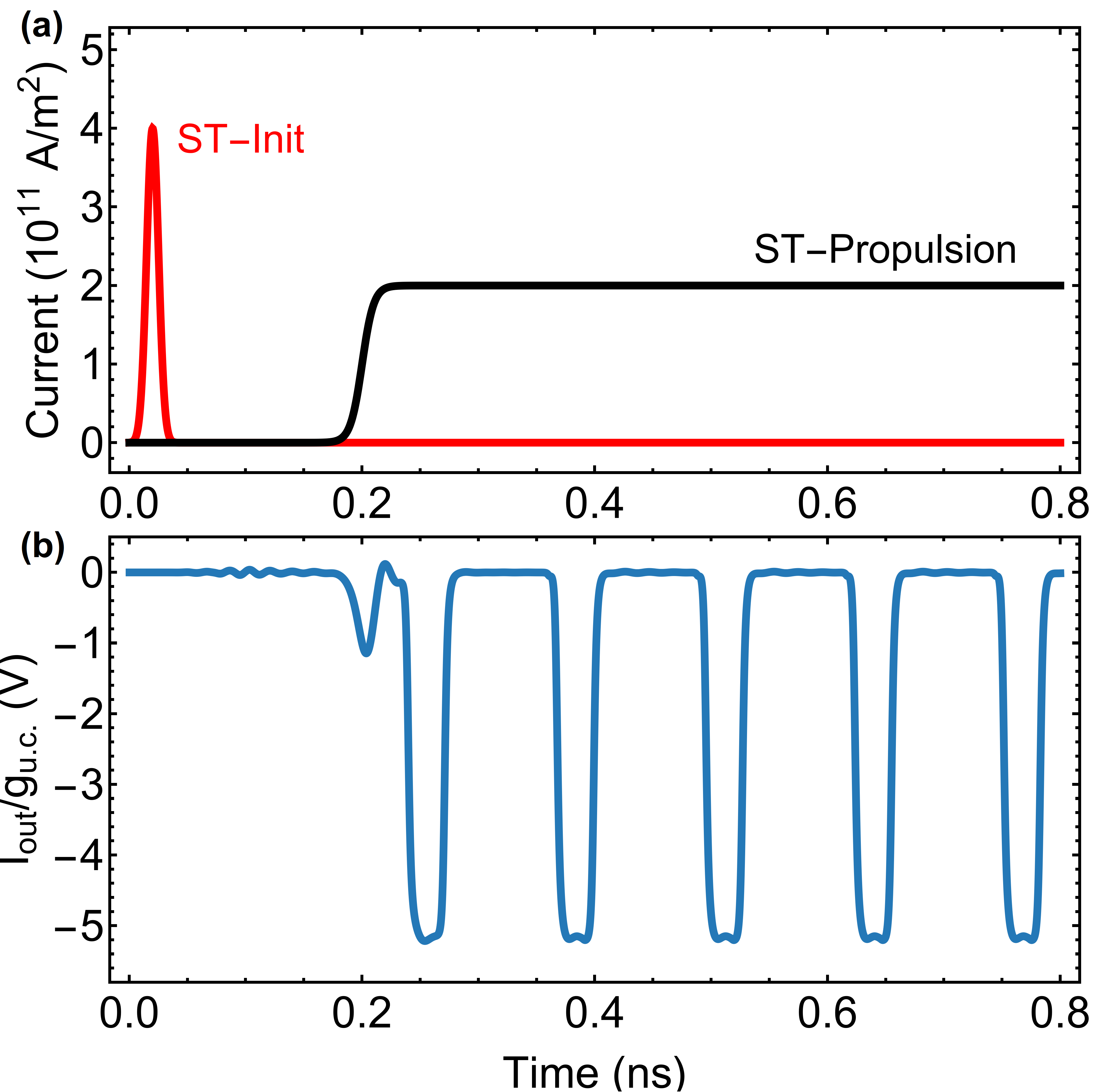}
\caption{The time dependence of (a) current densities on ST-Init and ST-Propulsion sources, and (b) spin accumulation in the centrally localized readout region, which spans a length of $L/4$. Here, $L=1.39$~$\mu$m is the length of the Cr$_2$O$_3$ sample; selected damping rate $\alpha = 1\cdot 10^{-3}$.}
\label{img:localizedReadout}
\end{figure}

The experimental study of the dependence (\ref{eq:velocity}) can be done by measuring the response produced by soliton motion. In an LJJ system of the characteristic size $L$, the periodic movement of the fluxon is associated with the average voltage $\langle V\rangle = v \Phi_0/L$  \cite{fulton_single_1973, levring_fluxon_1982}. The current-voltage characteristic thus reproduces the dependence of the fluxon velocity on the driving force, manifested in the so-called zero-field steps  \cite{fulton_single_1973, levring_fluxon_1982, ustinov_long_1998, ustinov_solitons_1998}. While this indirect measurement suggests various applications \cite{pedersen_fluxon_1991}, a more interesting problem is the possibility of direct localized detection, such as for a qubit state readout \cite{fedorov_fluxon_2014, fedorov_reading_2007, herr_kidiyarova-shevchenko_design_2007, averin_rapid_2006}.

Equivalent measurement schemes can be implemented to detect a BL moving within the DW. The spin dynamics caused by the BL movement induces nonzero output torque $\tau_{out}$, which is proportional to the projection of $\mathbf{l} \times \dot{\mathbf{l}}$, that has a sense of dynamical response here due to the spin pumping mechanism \cite{tserkovnyak_enhanced_2002, sinova_spin_2015}. Its sum over a specified region $S$ determines the spin voltage $V = (\hbar / e) \sum_S \tau_{out}$ that impinges on the interface and leads to the spin current $I_{out}$ flowing out of the AFM material; $I_{out} = V g_{u.c.}$, where $g_{u.c.}$ is the spin-mixing conductance on a unit cell. Thus, the local detection of the passage of the BL can be done by applying an additional localized layer of platinum, see the schematic setup in Fig. \ref{img:schema}. The generated spin current in this area will be read out as an electric one due to the inverse spin-Hall effect.

The temporal dependence of the output signal is determined by the convolution of the detector profile and the moving BL, see Fig. \ref{img:localizedReadout}. However, it is important to note that the detector also captures other dynamic phenomena. For instance, spin waves can be excited when the frequencies exceed the intrawall characteristic frequency, such as during the creation and reflection of the BL at the sample boundary. These spin waves cause small ripples in the output signal, but in the considered operational setup, they propagate faster than the BL, decay quickly, and do not impact the BL's dynamics. Another response to consider is the phase tilting of the DW, which occurs when the ST-Propulsion source is activated but has no further consequences except BL profile modification.

\section{Conclusions}

We show that in bi-axial antiferromagnets, the injection of spin currents into the DW is an effective way to create and move a BL. The dynamic state of this BL generates a time-varying torque, which can be detected through the inverse spin-Hall effect. The dynamics of such a BL is analogous to a fluxon in an LJJ, where injected spin torque act as an electric current between two superconductors. In contrast to an LJJ, the damping rate here can reach substantially high values and be of the same order as the intrawall characteristic frequency. The description of the BL dynamics, in this case, requires accounting for the structural changes of a DW caused by the spin current.

The BL spin texture is not the only magnetic analog to Josephson systems \cite{hill_spin-torque-biased_2018, khymyn_antiferromagnetic_2017}. However, it has the potential to be utilized in LJJ-inspired applications due to the additional degree of freedom provided by the DW that hosts a BL. A DW can serve as a reconfigurable transmission line that can be placed at a specific location within the device using a perpendicularly polarized spin current \cite{ovcharov_spin_2022, hartmann_nonlocal_2021}.

\section{Acknowledgements}
This project is partly funded by the European Research Council (ERC) under the European Union’s Horizon
2020 research and innovation programme (Grant TOPSPIN No 835068) and the Swedish Research Council Framework Grant Dnr. 2016-05980. B.A.I. acknowledges support from the National Research Foundation of Ukraine, under Grant No. 2020.02/0261.

\bibliography{main}

\end{document}